\documentclass[manuscript]{aastex}

\slugcomment{Article for The Astronomical Journal}

\shorttitle{An interpretation of Stephan's Quintet}
\shortauthors{Gibson and Schild}

\begin{document}


\title{Interpretation of the Stephan Quintet Galaxy Cluster using
Hydro-Gravitational Theory}


\author{Carl H. Gibson\altaffilmark{1}}
\affil{Departments of Mechanical and Aerospace Engineering and 
Scripps Institution
of Oceanography, University of
California,
     San Diego, CA 92093-0411}

\email{cgibson@ucsd.edu}

\and

\author{Rudolph E. Schild}
\affil{Center for Astrophysics,
     60 Garden Street, Cambridge, MA 02138}
\email{rschild@cfa.harvard.edu}


\altaffiltext{1}{Center for Astrophysics and Space Sciences, UCSD}


\begin{abstract}
Stephan's Quintet (SQ) is a compact group of galaxies that has been well
studied since its discovery in 1877 but is mysterious using cold
dark matter hierarchical clustering cosmology  (CDMHCC). Anomalous red shifts $z
= (0.0027,0.019, 0.022,  0.022,  0.022)$ among galaxies in SQ either; reduce it
to a Trio with two highly improbable intruders from CDMHCC, or support the
Arp (1973) hypothesis that its red shifts may be intrinsic.  An alternative is
provided by the Gibson 1996-2000 hydro-gravitational-theory (HGT) where
superclusters, clusters and galaxies all originate by universe expansion and
gravitational fragmentation in the super-viscous plasma epoch (after which the
gas condenses as $10^{24}$ kg fog-particles in metastable $10^{36}$
kg dark-matter-clumps).  By this fluid mechanical cosmology, the SQ galaxies
gently separated recently and  remain precisely along a line of sight because of
perspective and the small transverse velocities permitted by their sticky,
viscous-gravitational, beginnings.  Star and gas bridges and 
young-globular-star-cluster (YGC) trails  observed by the HST are triggered as
SQ galaxies separate through each other's frozen baryonic-dark-matter halos  of
dark proto-globular-cluster (PGC) clumps of planetary-mass
primordial-fog-particles (PFPs). Discordant red shifts (from CDMHCC) between
angularly clustered quasars and bright galaxies  are similarly explained by HGT.
\end{abstract}




\section{Introduction}

Stephan's Quintet (SQ, HGC 92, Arp 319, VV 288) is one of the first known
(Stephan 1877) and best studied of the Hickson 1982 catalog of very compact
groups of galaxies, and historically the most mysterious.  The group consists
of the Trio NGC 7319, NGC 7318A, and NGC 7317, all of which have redshift
0.022, NGC 7318B with redshift 0.019 closely aligned with NGC 7318A, and NGC
7320.  Burbidge and Burbidge 1959 noted that the large discrepancy of redshifts
for the double galaxy NGC 7318AB requires huge mass/light ($M/L$) ratios
$\approx 300 \pm 200$ from dynamical models to achieve virial equilibrium. 
However, the true mystery of SQ began when the missing redshift for NGC 7320
 was determined by Burbidge and Burbidge 1961 to be only $z = 0.0027$, with
relative velocity $cz = 8.1 \times 10^{5}$ m/s compared to $  6.7 \times
10^{6}$ for the Trio.  For virial equilibrium, this increases the kinetic
energy of the group by a factor 
$\sim 30$ and would require $M/L \approx 10,000$: much too large to be
credible.  Thus it was concluded
\citep{bur61} that the system is in a state of explosive expansion since
the $\it a$ $ \it priori$ chance of NGC 7320 not being a member of the group but
a random foreground galaxy  is about 1/1500.  

Another possibility is that the SQ red shifts are intrinsically variable because
the SQ galaxies were all recently ejected from a nearby parent AGN.   Arp 1973
summarizes several papers from 1970-1972 where he concludes that the nearby
large spiral galaxy NGC 7331 has ejected all the SQ galaxies and some have
intrinsic red shifts, so that all the SQ galaxies are located at the same
$\approx 10$ Mpc distance of their parent NGC 7331.  Arp has noted numerous
cases where galaxies in close angular proximity have not only widely different
red shifts but coincident spin magnitudes and alignments with the AGN jets,
consistent with his hypothesis that galaxies and quasars can be ejected from
active galactic nuclei (AGNs) with intrinsic red shifts
\citep{arp98}. 

Galaxies and quasars frequently show evidence of ejection with intrinsic red
shifts
\citep{hoy00}.  A Seyfert 1 galaxy (NGC 6212) is observed closely surrounded by
a large number  ($\ge$44) of QSOs that it may have ejected \citep{bur03}, with
QSO surface densities (69 per square degree) larger than ambient by estimated
factors of 30 to 10 and  decreasing  (to 17) with angular
distance for radii 10$\--$50 minutes.   It has  been suggested
\citep{hoy00} that the big bang hypothesis itself may be questioned based on
the remarkable accumulation of such coincidences that are contrary to the
statistics of standard CDM hierarchical galaxy clustering cosmology (CDMHCC)
and the Hubble red-shift  radial-velocity relationship ($v = cz$) of big bang
cosmology.  

The contradictions and
mysteries  vanish concerning SQ anomalous red-shifts and large
QSO densities near AGNs when the observations are interpreted using the
hydro-gravitational-theory (HGT) of Gibson 1996-2000.  From HGT cosmology such
improbable coincidences (from CDMHCC) are simply fossil manifestations of the
viscous-gravitational beginnings of galaxy clusters and galaxies, where the
apparent close proximity is an optical illusion resulting from the small
transverse velocities of the galaxies due to early friction as they gently
fragmented gravitationally from the same proto-supercluster. Thus, dense angular
galaxy clusters can have wide spatial separations precisely along a line of
sight from the uniform expansion of space expected from big bang cosmology.
From the large range of redshifts (0.03 to 2.6) and Hubble distances (150 to
3730 Mpc) of the NGC 6212-quasar system
\citep{bur03} the observed
AGN-QSO galaxies are concentrated in a
thin ($\approx$150/1) line-of-sight pencil, contradicting CDMHCC and
supporting HGT.  HGT has recently been reviewed and compared to data
\citep{gs03}, so in the
following we present only a brief summary.

\section{Hydro-Gravitational Theory}

Standard CDMHC cosmologies are based on over-simplified fluid
mechanical equations, an inappropriate assumption that the fluid is
collisionless, and a recognized ``swindle'' required to achieve solution of the
equations.  The Jeans 1902 theory neglects viscous forces, turbulence forces,
non-acoustic density fluctuations, particle collisions, and the effects of
diffusion on gravitational structure formation.  Jeans did linear perturbation
stability analysis (neglecting turbulence) of Euler's equations (neglecting
viscous forces) for a
nearly uniform ideal gas with density
$\rho$ only a function of pressure (the barotropic assumption),  which reduced
the problem of gravitational instability to the solvable equation of
gravitational acoustics.  To reconcile his equations with the linearized
collisionless Boltzmann's equations and the resulting Poisson's equation for the
gravitational potential, Jeans assumed the density $\rho$ was zero.  This
assumption is appropriately known as the ``Jeans swindle''.    The only critical
wave length for gravitational stability with all these questionable assumptions
is the Jeans length scale
$L_J$ where
\begin{equation}  L_J \equiv V_S/(\rho
G)^{1/2}
\approx (p/\rho^2 G)^{1/2} ,
\label{eq1}
\end{equation}
  $G$ is Newton's
gravitational constant and
$V_S \approx (p/\rho)^{1/2}$ is the sound speed.

Density fluctuations in fluids are not barotropic 
as assumed by
Jeans 1902 except rarely in small regions for short times near powerful sound
sources.  Density fluctuations that triggered the first gravitational
structures in the primordial fluids of interest were likely
non-acoustic (non-barotropic) density variations from turbulent mixing of
temperature or chemical species  concentrations produced by the big bang
\citep{gib01} as shown by turbulent signatures in the cosmic microwave
background temperature anisotropies \citep{bs02}.  From Jeans' theory without
Jeans' swindle, a gravitational condensation on an acoustical density maximum
rapidly becomes a non-acoustical density maximum  because the
gravitationally accreted mass retains the (zero) momentum of the
motionless ambient gas.

Without viscous and turbulent forces or diffusion, fluids with non-acoustic
density fluctuations are absolutely unstable to the formation of structure due
to self gravity
\citep{gib96}.  Turbulence or viscous forces can dominate gravitational forces
at small distances from a point of maximum or minimum density to prevent
gravitational structure formation, but gravitational forces will dominate
turbulent or viscous forces at larger distances to cause structures if the gas
or plasma does not diffuse away faster than it can condense or rarify due to
gravity.  The concepts of pressure support and thermal support are artifacts of
the erroneous Jeans criterion for gravitational instability.  Pressure forces
cannot prevent gravitational structure formation in the plasma epoch because
pressures equilibrate in time periods smaller that the gravitational free fall
time $(\rho G)^{-1/2}$ on length scales smaller than the Jeans scale, and
the Jeans scale in the primordial plasma is larger than the Hubble
scale of causal connection $L_H = ct$, where $c$ is light speed and $t$ is
time.  Therefore, if gravitational forces exceed viscous and turbulence forces
in the plasma epoch at scales smaller than $L_H$ then gravitational structures
will develop, independent of the Jeans criterion.  Only a very large
diffusivity of the plasma ($D \gg \nu$) could interfere.

The
diffusion velocity is
$D/L$ for diffusivity
$D$ at distance $L$ and the gravitational velocity is $L \rho^{1/2} G^{1/2}$.
The two velocities are equal at the diffusive Schwarz length scale
\begin{equation} L_{SD} \equiv
[D^2 / \rho G]^{1/4}.\end{equation}  Thus very weakly collisional particles such
as the hypothetical cold-dark-matter (CDM) material cannot form 
potential wells
for baryonic matter collection because the particles have large diffusivity
and will disperse, consistent with observations \citep{sa02}.  Diffusivity
$D 
\approx V_p \times
L_c$, where $V_p$ is the particle speed and $L_c$ is the collision distance.
Because weakly collisional particles have large collision distances with
large diffusive Schwarz lengths the non-baryonic dark matter
(possibly neutrinos) is the last material to fragment by self gravity and
not the first as assumed by CDM cosmologies.  The first structures occur as
proto-supercluster-voids in the baryonic plasma controlled by viscous and weak
turbulence forces, independent of diffusivity  ($D \approx \nu$).  The CDM
seeds postulated as the basis of CDMHCC never happened because $(L_{SD})_{NB}
\gg ct$ in the plasma epoch.

The baryonic matter is subject to large viscous
forces, especially in the hot primordial plasma and gas states 
existing when most
gravitational structures first formed.  The viscous forces per unit 
volume $\rho
\nu
\gamma L^2$ dominate gravitational forces $\rho^2 G L^4$ at small scales, where
$\nu$ is the kinematic viscosity and $\gamma$ is the rate of strain of the
fluid.  The forces match at the viscous Schwarz length
\begin{equation} L_{SV} \equiv (\nu \gamma /
\rho G)^{1/2},\end{equation}
  which is the smallest size for self gravitational condensation or
void formation in such a flow.  Turbulent forces may require even larger scales
of gravitational structures.  Turbulent forces $\rho \varepsilon^{2/3} L^{8/3}$
match gravitational forces at the turbulent Schwarz scale
\begin{equation}L_{ST} \equiv \varepsilon
^{1/2}/(\rho G)^{3/4},\end{equation}
  where $\varepsilon$ is the viscous dissipation rate of the turbulence. 
Because in the primordial plasma the viscosity and diffusivity are identical
and the rate-of-strain $\gamma$ is larger than the free-fall frequency $(\rho
G)^{1/2}$, the viscous and turbulent Schwarz scales
$L_{SV}$ and 
$L_{ST}$ will be larger than the diffusive Schwarz scale $L_{SD}$, from
(2), (3) and (4).  

Therefore, the criterion for structure formation in the plasma epoch is that
both $L_{SV}$ and $L_{ST}$ become less than the horizon scale $L_H = ct$. 
Reynolds numbers in the plasma epoch were near critical, with  $L_{SV} \approx
L_{ST}$.  From $L_{SV}< ct$ and (3), gravitational structures first formed when
$\nu < c^2 t (t^2 \rho G) \approx c^2 t$ at time $t \approx 10^{12}$ seconds
\citep{gib96}, well before $10^{13}$ seconds which is the time of plasma to gas
transition (300,000 years).  Because the expansion of the universe inhibited
condensation but enhanced void formation in the weakly turbulent plasma, the
first structures were proto-supercluster-voids. At $10^{12}$ s 
\begin{equation}
(L_{SD})_{NB} \gg L_{SV} \approx L_{ST} \approx 5 \times L_K \approx L_H = 3
\times 10^{20} \rm m, \end{equation} 
where $L_{SD}$ applies to the non-baryonic component and $L_{SV}$, $L_{ST}$,
and $L_{K}$ apply to the
baryonic component.

As proto-supercluster fragments formed the voids filled with
non-baryonic matter by diffusion, inhibiting further structure formation by
decreasing the gravitational driving force.  The baryonic mass density
$\rho
\approx 2
\times  10^{-17}$ kg/$\rm
m^3$ and rate of strain
$  \gamma \approx 10^{-12}$ $\rm s^{-1}$ were preserved as hydrodynamic fossils
within the proto-supercluster fragments and within
proto-cluster and proto-galaxy objects resulting from subsequent fragmentation
as the photon viscosity and
$L_{SV}$ decreased prior to the plasma-gas transition and photon decoupling
\citep{gib00}.  As shown in Eq. 5, the Kolmogorov scale $L_K \equiv [\nu^3
/\varepsilon ]^{1/4}$ and the viscous and turbulent Schwarz
scales at the time of first structure nearly matched the
horizon scale $L_H
\equiv ct \approx 3 \times 10^{20}$ m, freezing in the density, strain-rate, and spin magnitudes and
directions of the subsequent proto-cluster and proto-galaxy fragments of
proto-superclusters.  Remnants of the strain-rate and spin magnitudes and
directions of the weak turbulence at the time of first structure formation are
forms of fossil vorticity turbulence
\citep{gib99}. Thus, HGT explains galaxy spin
alignments and close angular associations with quasars 
without assuming intrinsic red shifts and mutual ejections.

The quiet condition of the primordial gas is  revealed by
measurements of temperature fluctuations of the cosmic microwave background 
radiation that show
an average $\delta T/T \approx 10^{-5}$ much too small for any 
turbulence to have
existed at that time of plasma-gas transition ($10^{13}$ s).  Turbulent plasma
motions are strongly damped by buoyancy forces at horizon scales after the first
gravitational fragmentation time 
$10^{12}$ s.  Viscous forces in the plasma are inadequate to explain the lack
of primordial turbulence ($\nu$
$ \ge 10^{30}$ m$^2$ s$^{-1}$ is required but, after $10^{12}$ s, $\nu \le 4
\times 10^{26}$, Gibson 2000). Thus the observed lack of plasma turbulence
proves that large scale buoyancy forces and gravitational structure formation
must have begun in the plasma epoch.

The gas temperature,
density, viscosity, and rate of strain are all precisely known at transition,
so the gas viscous Schwarz  mass
$L_{SV}^3 \rho$ is easily calculated to be about
$10^{24}$ kg, the mass of a small planet, or about
$10^{-6} M_{\sun}$, with uncertainty about a factor of ten.  From HGT, soon
after the cooling primordial plasma turned to gas at $10^{13}$ s (300,000 yr),
the entire baryonic universe condensed to a fog of planetary-mass
primordial-fog-particles (PFPs).  These gas-cloud objects gradually cooled,
formed H-He rain, and eventually froze solid to become the baryonic dark
matter and the basic material of  construction for stars and everything else,
about $30 \times 10^{6}$ rogue planets per star.

The Jeans mass $L_J^3 \rho$ of the primordial gas at transition was about $10^6
M_{\sun}$ with about a factor of ten uncertainty, the mass of a globular star
cluster.  Proto-galaxies fragmented at the PFP scale but also at this
proto-globular-star-cluster PGC scale
$L_J$, although
not for the reason given by the Jeans 1902 theory.  Density fluctuations in the
gaseous proto-galaxies were absolutely unstable to 
void formation
at all scales larger than the viscous Schwarz scale $L_{SV}$.  Pressure can 
only remain in
equilibrium with density without temperature changes in a gravitationally
expanding void on scales smaller than the Jeans scale.  From the second law of
thermodynamics, rarefaction wave speeds that develop as density 
minima expand due
to gravity to form voids are limited to speeds less than the sonic velocity. 
Cooling would therefore occur and be compensated by radiation in the otherwise
isothermal primordial gas when the expanding voids approached the Jeans scale.
Gravitational fragmentation of  proto-galaxies will then
be accelerated by radiative heat transfer to these cooler regions, resulting in
fragmentation at the Jeans scale and isolation of proto-globular-star-clusters
(PGCs) with the primordial-gas-Jeans-mass.  

These
$10^{36}$ kg PGC objects were not able to collapse from their own self
gravity  because of their
internal fragmentation at the viscous Schwarz scale to form $10^{24}$ kg PFPs.
The fact that globular star clusters have precisely the same density and
primordial-gas-Jeans-mass from galaxy to galaxy proves they were all formed
simultaneously soon after the time of the plasma to gas
transition $10^{13}$ s.  The gas has never been so uniform since, and no
mechanism exists to recover such a high density, let alone such a high uniform
density, as the fossil turbulent density value $\rho \approx 2 \times 10^{-17}$
kg/$\rm m^3$.  Young globular cluster formation in BDM halos in the Tadpole,
Mice, and Antennae galaxy mergers show that dark PGC clusters of PFPs are
remarkably stable structures, persisting without disruption or star formation
for more than ten billion years.

\section{Stephan's Quintet:  HGT interpretation of HST image}

Moles et al. 1997 summarize the data and dynamical status of SQ consistent with
standard CDMHC cosmology, proposing that the nearby NGC 7320C with $cz = 6.0
\times 10^{5}$ m/s (matching that of NGC 7318B) has possibly collided several
times with SQ members stripping their gas and central stars to form luminous
wakes and to  preserve their dynamical equilibrium, thus accounting for the
fact that 43 of the 100 members of the Hickson 1982 catalog of compact groups
contain discordant redshift members.  However, Gallagher et al. 2001 show from
their Hubble Space Telescope (HST) measurements that globular star clusters in
SQ are not concentrated in the inner regions of the galaxies as observed in
numerous merger remnants, but are spread over the SQ debris and surrounding
area.  We see no evidence of collisions or mergers in the HST images of SQ and
suggest the luminous wakes are not gas stripped from galaxy cores by collisions
but are new stars triggered into formation in the baryonic-dark-matter halo of
the SQ cluster as member galaxies are gently stretched away by the expansion of
space.  

According to HGT, galaxy mergers and collisions do not strip gas but
produce gas by evaporating the frozen hydrogen and helium of the planetary mass
objects which dominate the baryonic mass of galaxies.  The baryonic dark matter
is comprised of proto-globular-star-cluster (PGC) clumps of planetary-mass
primordial-fog-particles (PFPs) from hydro-gravitational-theory 
\citep{gib96} and quasar microlensing observations \citep{sch96}. Therefore the
cores of SQ galaxies should be deficient in gas and YGCs because they have not
had mergers or collisions.  

Following standard CDMHC cosmology and N-body computer models,
galaxies and clusters of galaxies are formed by hierarchical collisionless
clustering due to gravity starting with
sub-galaxy mass CDM seeds condensed in the plasma epoch after the big bang.  The
Jeans 1902 gravitational condensation criterion rules out structures forming in
ordinary baryonic matter.  CDM seeds are diffusionally unstable from
hydro-gravitational theory and their clustering to form galaxies is contrary to
observations
\citep{sa02}.  From HGT, both CDMHC
cosmology and the Jeans 1902 criterion are fundamentally incorrect and
misleading
\citep{gib00}.  The unknown non-baryonic CDM material is enormously diffusive
compared to the H and He ions of the primordial plasma and cannot condense or
fragment gravitationally.  However, we can be sure structure formation
occurred in the plasma epoch because buoyancy within self
gravitational structure is the only  mechanism available to prevent
turbulence.   Viscous forces were quite inadequate. 
Fully developed turbulence would have produced $\delta T/T
\approx 0.1$ values much larger than the $\delta T/T \approx 0.00001$ values
observed in numerous cosmic microwave background studies.  From HGT, structure formation first
occurred by gravitational fragmentation due to the expansion of space when
viscous and weak turbulence forces of the primordial plasma matched
gravitational forces at scales smaller than the horizon scale
$ct$, where $c$ is the speed of light and $t$ is the time after the big bang. 
The growth of structure was arrested by non-baryonic matter filling the voids
between baryonic fragments.  This HGT-cosmology and its application to
the interpretation of SQ is illustrated schematically in Figure 1ab.

In Fig. 1a at top left we see a fragmenting proto-supercluster ($10^{47}$ kg) of
the primordial plasma as it separates from other such fragments due to the
rapid expansion of the universe at the time of first gravitational structure
formation about 30,000 years after the big bang
\citep{gib96}.  The scale is near the horizon scale $ct$ at that time $3 \times
10^{20}$ m with baryonic density $2 \times 10^{-17}$ kg/$\rm m^3$ and
non-baryonic density $
\approx 10^{-15}$ kg/$\rm m^3$  decreasing with time
and the non-baryonic matter (possibly neutrinos) diffuses to fill the voids and
reduce the gravitational forces \citep{gib00}.  In Fig. 1a center proto-cluster
fragments form and separate, and on the right
proto-galaxies fragment just before the cooling plasma turns to gas at 300,000
years ($10^{13}$ s).  

The proto-galaxies preserve the density and spin of the
proto-supercluster as fossils of the primordial plasma turbulence
\citep{gib99}.  Their initial size is therefore about $5 \times 10^{19}$ m. 
These fragment into Jeans-mass  ($10^{36}$ kg) proto-globular-cluster (PGC)
dense clouds of ($10^{24}$ kg) primordial-fog-particles (PFPs) that cool,
freeze, and diffuse away from the galaxy cores to form baryonic-dark-matter
(BDM) halos around galaxies and galaxy-clusters such as SQ.  The Jeans-mass is
relevant, but not for the reasons given by Jeans (1902).  Some galaxy-clusters
can be very slow in their separation due to crowding and frictional forces of
their BDM halos, as shown by the central galaxy cluster at the right of Fig.
1a. The BDM halo may reveal the history of galaxy mergers and separations
because strong tidal forces and radiation by galaxy cores trigger the formation
of stars and YGCs as they and their  halos move through each other's BDM halos,
leaving star wakes and dust wakes.

Fig. 1b shows schematically our interpretation of SQ based on HGT.  The five
galaxies are separated by distances inferred from Hubble's law and their red
shifts times the horizon distance $10^{26}$ m due to the stretching of space
along a thin square tube of diameter $\approx 2
\times 10^{21}$ m oriented along the line of sight to the Trio.  The distance to
the line-of-sight tube entrance from earth is thus
$\approx 2.7
\times 10^{23}$ m for NGC 7320, with the exit and Trio at  $\approx 2.2
\times 10^{24}$ m.  NGC 7320 appears larger than the Trio members because it is
closer, consistent with the fact that it contains numerous obvious
young-globular-clusters (YGCs) from the HST images, but YGCs in the Trio are
barely resolved \citep{gal01}.  The tube in Fig. 1b is not to scale: the true
aspect ratio is that of a sheet of paper or a very long stick of uncooked
spaghetti.  By perspective,  about 1\% of the front face of the tube covers the
back face.

Figure 2 shows an HST image of Stephan's Quintet.  The trail of luminous
material extending southeast of NGC 7319 is interpreted from HGT as a star wake
formed as one of the galaxy-fragments of the original cluster moves away
through the baryonic-dark-matter (BDM) halo, triggering star formation until it
exits at the halo boundary marked by a dashed line.  Other star wakes in Fig. 2
are also marked by arrows.  These star wakes are similar in origin to the
filamentary galaxy VV29B of the Tadpole merger
\citep{gs03} and the ``tidal tails'' of the Mice and Antennae merging galaxies,
except that in SQ all the galaxies are seen to separate through each
other's halos rather than merge, contrary to the standard SQ \citep{mol97}
model.  

Two dust trails
are shown by arrows in the upper right of Fig. 2 that we interpret as star
wakes of the separation of NGC 7318B from NGC 7318A.  A similar dust trail is
interpreted from its direction as a star wake of NGC 7331 produced in the NGC
7319 BDM halo as it moved out of the cluster.  The luminous trail pointing
toward NGC 7320C is confirmed by  gas patterns \citep{gut02} observed from
broadband R measurements that suggest NGC 7320 has the same origin near NGC
7319.  An unidentified galaxy separated in the northern star
forming region, leaving over a hundred YGCs \citep{gal01} before  exiting the
BDM halo boundary (shown by the dashed line in the upper left of Fig. 2).

Details of the Hubble Space Telescope images of Stephan's Quintet (including
Fig. 2) can be found at the website for the July 19, 2001 STScI-2001-22 press
release (http://hubblesite.org/newscenter/archive/2001/22/image/a).  The images
are described as ``Star Clusters Born in the Wreckage of Cosmic Collisions''
reflecting the large number of YGCs detected \citep{gal01} and the standard SQ
model \citep{mol97}.  

According to our HGT interpretation, none of the YGCs are
due to galaxy collisions or mergers.  All are formed in the BDM halos as the
galaxies gently separate with small transverse velocity along lines of
sight.  There were no cosmic collisions and there is no wreckage.  Numerous
very well resolved YGCs can be seen in the NGC 7320 high resolution image with
separations indicating numbers in the range
$10^{5} \-- 10^{6} $.  This suggests a significant fraction the dark baryonic
matter in the halo of NGC 7320 has been triggered to form YGCs and stars as the
galaxy separated through both the dense BDM halo of the SQ Trio and the BDM
halo of its companion galaxy NGC 7331, also at $z=0.0027$.   No such
concentration of YGCs can be seen in the SQ Trio galaxies, consistent with our
HGT interpretation that they are at 8.3 times the distance of NGC 7320 as shown
in Fig. 1b.

\section{Stephan's Quintet:  HGT interpretation of R and $H_\alpha$ maps}

The present status of observations of Stephan's Quintet is well summarized 
by Gutierrez et al. 2002, including their deep broadband R and narrowband
$H_\alpha$ maps shown in Figure 3.  The R band map (their Fig. 1) with
sensitivity 26 mag arcsec$^{-2}$ extends to a wide range that includes NGC
7320C with the other SQ member galaxies.  A clear $H_\alpha$ bridge is shown
with red shift
$z=0.022$ corresponding to that of the SQ Trio to a sharp interface with
$z=0.0027$ material in NGC 7320, consistent with our interpretation that the
bridge was formed in the BDM halo of the SQ Trio by NGC 7320 as it emerged and
separated by the expansion of the universe along the line of sight, as shown by
the dashed arrow in Fig. 2.  

A corresponding dashed arrow in Fig. 3 shows the new $H_\alpha$ bridge from the
SQ Trio at red shift 0.022 with sharp transition to 0.0027 at NGC 7320, proving
the two are widely separated in space but likely with the same origin, as we
show is expected from HGT in Fig. 1b.

The solid arrow shown in Fig. 3 toward NGC 7320C suggests its emergence from
the SQ Trio BDM halo leaving the star wake shown by a corresponding arrow in
Fig. 2.  The mechanism of star wake production is that the frozen PFPs are in
meta-stable equilibrium within their PGCs.  Radiation from a passing galaxy 
causes evaporation of gas and tidal forces which together  increase the rate of
accretion of the PFPs to form larger planets and finally stars.  The size of
the stars and their lifetimes depends on the turbulence levels produced in the
gas according to HGT.  Large turbulence levels produce large, short lived
stars.  The dust lane between NGC 7318A and its double NGC 7318B suggests large
turbulence levels produced large stars that have since turned to dust through
supernovas. A similar dust lane from NGC 7219 is in the general direction of
NGC 7331 and its companions, as indicated by the arrow in Fig. 2.

\section{Conclusions}

We conclude that Stephan's Quintet is well described by
hydro-gravitational-theory and cosmology \citep{gib96}. According to HGT
cosmology, all the SQ galaxies formed in a cluster by gravitational
fragmentation of the primordial plasma just before photon decoupling and
transition to gas 300,000 years after the big bang.  None of the galaxies show
evidence of subsequent collisions or mergers.  They remained stuck together for
12.9 billion years until 220 million years ago when the uniform expansion of
space in the universe finally overcame gravitational forces and
frictional forces of the cluster baryonic-dark-matter halo.

The nature of the baryonic-dark-matter halo is explained by HGT and supported by
the SQ observations.  At the plasma-gas transition the proto-galaxy plasma
clouds turned to gas. From HGT \citep{gib96} the gas  fragmented at both the
Jeans scale, to form proto-globular-star-cluster (PGC) clumps ($10^{36}$ kg),
and the viscous Schwarz scale,
 to form small-planetary-mass ($10^{24}$ kg)
primordial-fog-particles (PFPs), consistent with the conclusion
\citep{sch96} from quasar microlensing observations that the lens galaxy mass
is dominated by ``rogue planets likely to be the missing mass''.  

Some of the
PFPs near the proto-galaxy centers accreted to form stars and the luminous
galaxy cores. Most PFPs condensed and froze as the universe expanded and cooled
so their PGCs remained dark and gradually diffused away from the galaxy cores
to form BDM galaxy halos, and some diffused further to form cluster
baryonic-dark-matter (BDM) halos.   The Stephan Quintet cluster BDM halo
boundaries are revealed by the separation of the SQ galaxies as star wakes, as
shown in Fig. 2.  The SQ BDM halo radius is only
$\approx 2
\times 10^{21}$ m,  compared with the BDM halo radius of the
Tadpole galaxy $\approx 5 \times 10^{21}$ m as shown
by HST/ACS images with the star wake of the merging galaxy \citep{gs03}.

Our HGT interpretation of SQ solves the long standing mystery of its anomalous
red shifts \citep{bur61}.  Rather than an explosive expansion or intrinsic red
shifts of the SQ galaxies ejected by the same parent \citep{arp73} we suggest
from HGT that a uniform expansion of the universe stretched the SQ
galaxies along a line of sight because of perspective and
small transverse velocities resulting from BDM halo gas friction and their
sticky beginnings,  as shown in Fig. 1b.  The common point of origin of the SQ
galaxies is confirmed by gas trails in recent R and
$H_\alpha$ maps, as shown in Fig. 3
\citep{gut02}. Discordant red shifts for
aligned quasars and AGN galaxies are also explained using HGT rather
than CDMHCC for the same reasons.

\acknowledgments

\clearpage

\begin{figure}
        \epsscale{0.6}
        \plotone{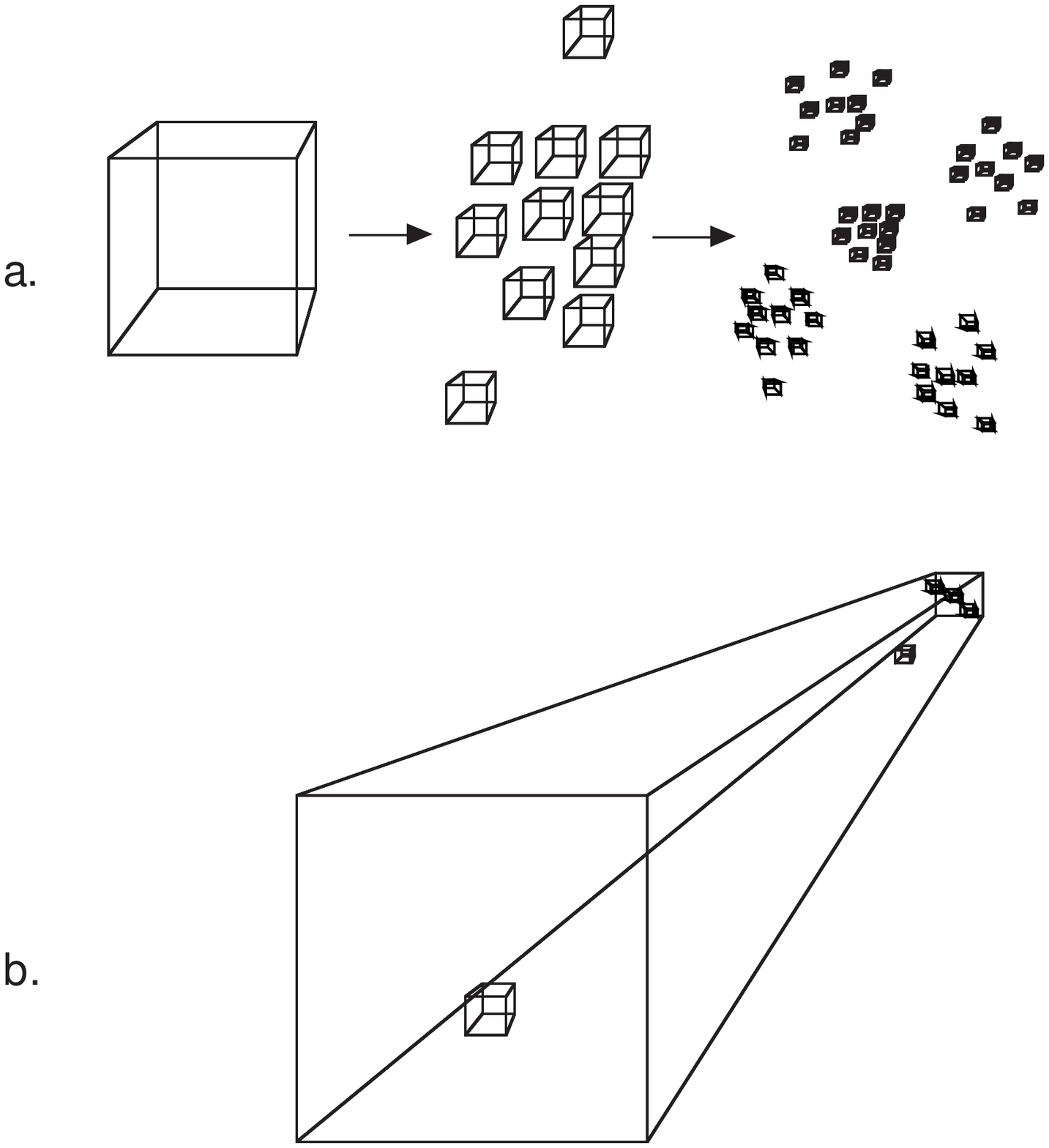}
        \caption{a.  According to hydro-gravitational cosmology
\citep{gib96}, proto-superclusters (left) fragment to proto-clusters (center)
which fragment to form proto-galaxies during the super-viscous plasma epoch. 
Compact galaxy clusters such as Stephan's Quintet occur in this cosmology
when dispersal of the cluster by the expansion of the universe is delayed by
frictional forces; eg., the central cluster of galaxies on the right.  b. 
Galaxies of the fragmented SQ cluster remain along a line of sight to the SQ
Trio because of their small transverse velocities, reflecting their sticky
beginnings.  The $2 \times 10^{21}$ m (60 kpc) diameter SQ thin tube begins
with NGC 7320 at a distance of
$2.7
\times 10^{23}$
 m (9 Mpc) and ends with the SQ Trio at $2.2 \times 10^{24}$ m (74 Mpc).  NGC
7318B is 10 Mpc closer than the Trio.}
       \end{figure}

\begin{figure}
        \epsscale{0.6}
        \plotone{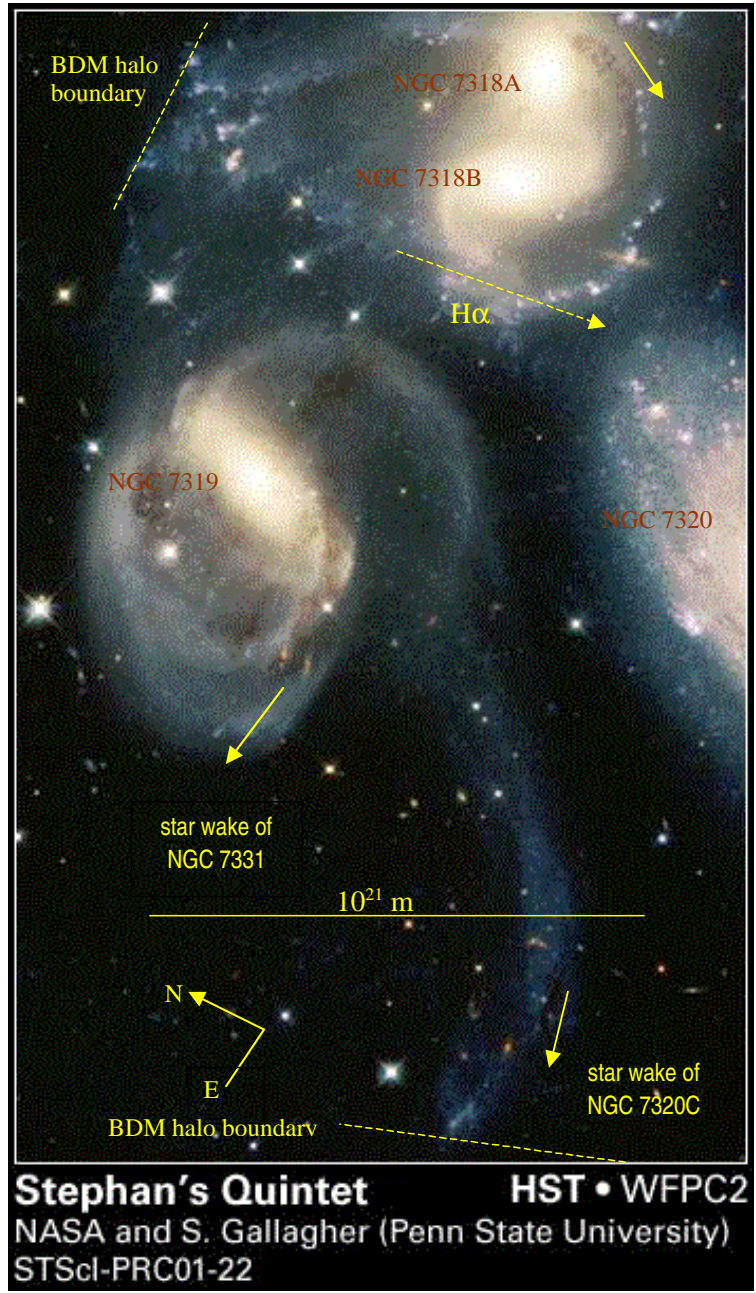}
        \caption{Hubble Space Telescope image of Stephan's Quintet. Dust and
star wakes (arrows) are produced as SQ related galaxies gently separate from
each other through the cluster baryonic-dark-matter (BDM) halo of PGCs and PFPs,
triggering star formation.  Star wakes of mergers and collisions are not
observed.}
       \end{figure}

\begin{figure}
        \epsscale{1.0}
        \plotone{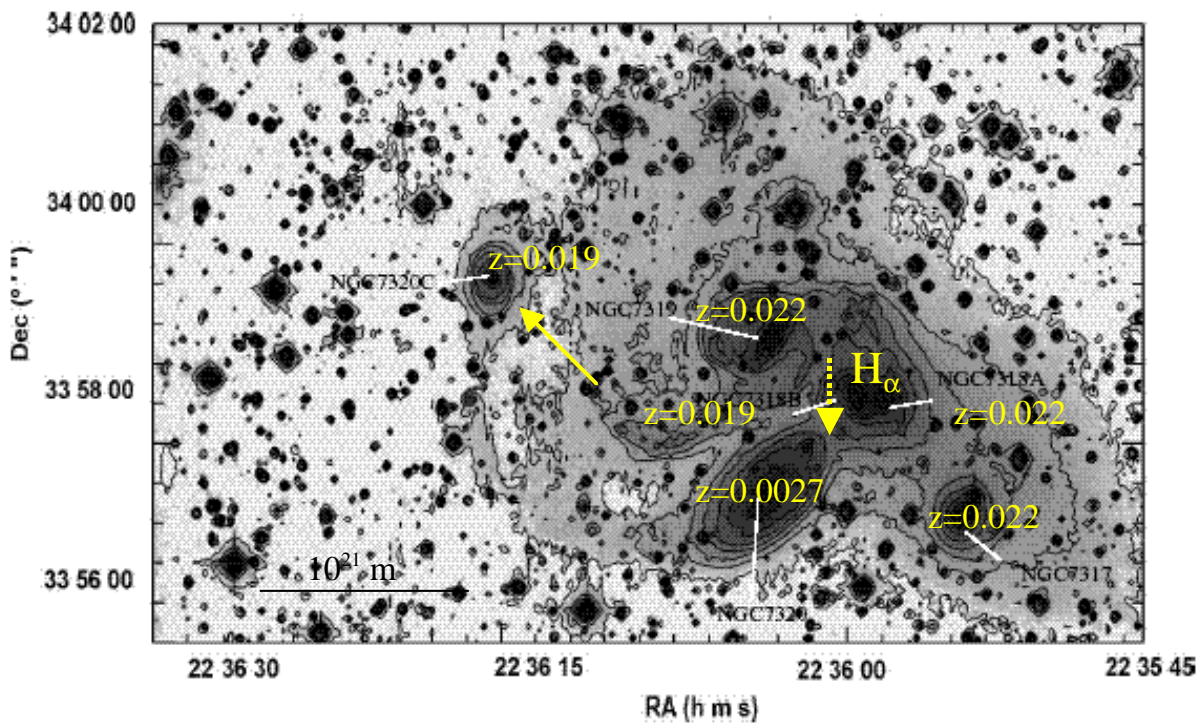}
        \caption{Contour R map of SQ \citep{gut02} showing connections between
SQ galaxies and NGC 7320C to the East (left), and NGC 7320 to the South
(bottom).  The $H_\alpha$ bridge is at the red shift 0.022 of the SQ Trio, and
shows a sharp transition to $z=0.0027$ for NGC 7320 \citep{gut02}, consistent
with our HGT interpretation that SQ galaxies have been stretched along a thin
pencil by the expansion of the universe, see Fig. 1b. }
       \end{figure}

\end{document}